\documentclass[]{lipics-iclp}
\lipicsheading{year}{numbers}{city}

\theoremstyle{plain}
\theoremstyle{definition}

\begin{document}

\title[Structured Interactive Scores]{Structured Interactive Musical Scores}

\author[lab1]{M. Toro}{Mauricio Toro-Berm\'{u}dez}
\address[lab1]{Universit\'{e} de Bordeaux 1, Laboratoire Bordelais de Recherche en Informatique, B\^{a}timent A30. 351, cours de la Lib\'{e}ration F-33405 Talence cedex, France.}  
\urladdr{http://www.labri.fr/perso/mtoro/}  


\keywords{ntcc, ccp, interactive scores, temporal relations, faust, ntccrt, heterogeneous systems, automatic verification}
\subjclass{D. Software, D.1 Programming techniques, D.1.3 Concurrent programming, D.1.5 Logic programming.}


\begin{abstract}
  \noindent 
Interactive Scores is a formalism for the design and
performance of interactive scenarios that provides temporal
relations (TRs)  among the objects of the scenario.
We can model TRs among objects in Time Stream Petri nets, but it is difficult to represent 
global constraints. This can be done
explicitly in the Non-deterministic Timed Concurrent Constraint (ntcc) calculus.
We want to formalize a heterogeneous system that controls in one subsystem
the concurrent execution of the objects using ntcc, and audio and video processing in the other.
We also plan to develop an automatic verifier for ntcc.
\end{abstract}

\maketitle

\section*{Introduction}\label{S:one}

\textit{Interactive Scores (IS)} are currently used for the design and performance of Electroacoustic
music \cite{aadc08} and live spectacles \cite{bamcrs09} (e.g., interactive theater plays and interactive museums). 
 Both applications
are based on Petri nets \cite{artech2008}.
The main purpose of IS is to provide temporal
relations; for instance, precedence between two objects and relations between their durations. Recently, we extended IS to support conditional branching together with temporal relations \cite{tbdcb10}. It is now possible to represent loops and choices.

We can model temporal relations in \textit{Time Stream Petri nets (TSPN)} \cite{SSW95}, but it is difficult to represent 
global constraints involving (possibly) all the objects of the scenario. Instead, in \textit{Concurrent Constraint Programming} (\texttt{ccp}) \cite{ccp} there are agents that reason about partial information contained in a constraint \textit{store}; thus,  global constraints are inherent in \texttt{ccp}. However, there is not discrete time in \texttt{ccp}, which makes it difficult to represent  reactive systems.

There are some IS models
based on extensions of \texttt{ccp} with discrete time. An example is the \textit{Non-deterministic Timed Concurrent Constraint} (\texttt{ntcc}) model of IS \cite{ntcc, AADR06}. \texttt{Ntcc} is an extension of \texttt{ccp} for non-determinism, asynchrony and discrete time. 

Ntcc belongs to a family of formalisms called process calculi. Process calculi has been applied to the modeling of interactive multimedia systems
 \cite{toro2016faust, toro2016gelisp, 2016arXiv160202169T, toro2015ntccrt, is-chapter,tdcr14,ntccrt,cc-chapter,torophd,torobsc,Toro-Bermudez10,Toro15,ArandaAOPRTV09,tdcc12,toro-report09,tdc10,tdcb10,tororeport},  spatially-explicit ecological systems \cite{EPTCS2047, PT13,TPSK14,PTA13,mean-field-techreport}, data Structures \cite{PAT2016, MorenoPT17, RestrepoPT17}.

In the declarative view, \texttt{ntcc} processes can be interpreted as \textit{linear temporal logic} formulae \cite{Pnu77}. The \texttt{ntcc} model includes an inference system in this logic to verify properties of  \texttt{ntcc} models. This inference procedure was proved to be of exponential time complexity. Nevertheless, we believe practical automatic verification could be envisioned  for useful subsets of \texttt{ntcc} via model checking (see \cite{fv06}). At present, there is no such automatic verifier for \texttt{ntcc}.

Automated verification for \texttt{ntcc} will provide information about the correctness of the system to computer scientists, and will provide important properties about the scenario to its designers and users; for instance, reachability and liveness. We plan to augment \texttt{ntcc} models of IS with these features.
\section{Current and future work}

\textit{Functional AUdio STream (Faust)} \cite{faust} is a programming language
for signal processing with formal semantics and \textit{Ntccrt} \cite{ntccrt} is a real-time capable interpreter for \texttt{ntcc}.
We implemented a signal processing prototype where Faust and Ntccrt interact together.
In the future, we want to define formal semantics to describe a heterogeneous system that includes three subsystems: (i) one based on \texttt{ntcc} to control
discrete events from the user and to synchronize the objects of the scenario, (ii)
another one based on Faust to process audio and video, and finally (iii) one in charge
to load and play audio and video files. 

At the time of this writing, there are no formal semantics of a heterogeneous system that synchronizes concurrent objects, handles global constraints, and controls audio and video streams. Modeling this kind of systems will be useful in other domains such as \textit{machine musical improvisation} \cite{Assayag-using} and \textit{music video games}. An advantage over the existing implementations of these systems will be verification.

In the proof system of \texttt{ntcc}, we can prove properties like  
``10 time units (TUs) after the event $e_A$, during the next 4 TUs, the stream $B$ is the result of applying a \textit{gain filter} to the stream $A$''. 
However, real-time audio processing cannot be implemented in Ntccrt because it requires to simulate 44100 TUs per second to process a 44.1 kHz sound. If we replace some \texttt{ntcc} processes by Faust plugins, we can execute such system efficiently, but we cannot verify that the properties of the system hold. 

There are two open issues: (i) how to 
prove that a Faust plugin that replaces a \texttt{ntcc} process respect the temporal properties proved for the process, and (ii) whether an implementation of Interactive Scores in Ntccrt can be as efficient as the existing Petri nets implementation, or as one using synchronous languages such as \textit{Signal} \cite{signal}, although the performance results from Ntccrt are promising\footnote{We ran a prototype of a score with conditional branching in Ntccrt. The score contains 500 temporal objects. The average duration of each time-unit was 30 ms, which is compatible with real-time interaction.}.

\section*{Acknowledgement}
I wish to acknowledge fruitful discussions with my supervisors Myriam Desainte-Catherine and Camilo Rueda.
I also want to thank them for helping me writing this article and for guiding my current research.

\bibliographystyle{lipics}	

\end{document}